\documentclass[12pt,twoside]{article}
\usepackage{wrapfig}
\usepackage{graphicx}
\usepackage{cmp2e}
\title[The effect of uniaxial crystal-field anisotropy \ldots ]%
{The effect of uniaxial crystal-field anisotropy on magnetic properties of the superexchange             antiferromagnetic Ising model}

\author[L. \v{C}anov\'a and M. Ja\v{s}\v{c}ur]{L. \v{C}anov\'a and M. Ja\v{s}\v{c}ur}

\address{Department of Theoretical Physics and Astrophysics, Faculty of Science, 
P.J.\v{S}af\'arik University, Park Angelinum 9, 040 01 Ko\v{s}ice, 
Slovak Republic}

\begin{document}
\maketitle

\begin{abstract}
The generalized Fisher super-exchange antiferromagnetic model with uniaxial crystal-field anisotropy is exactly investigated using an extended mapping technique. An exact relation between partition function of the studied system 
and that one of the standard zero-field spin-$1/2$ Ising model on the corresponding lattice is obtained applying the decoration-iteration transformation. Consequently, exact results for all physical quantities are derived for arbitrary spin values $S$ of decorating atoms. Particular attention is paid to the investigation of the effect of crystal-field anisotropy and external longitudinal magnetic field on magnetic properties of the system under investigation. The most interesting numerical results for ground-state and finite-temperature phase diagrams, thermal dependences of the sublattice magnetization and other thermodynamic quantities are discussed.

\pacs 75.10.Dg; 05.50.+q; 05.70.Jk; 75.50.Ee
\end{abstract}

\section{Introduction}
The Ising model turned out to be very convenient model for the investigation of magnetic properties of
strongly anisotropic materials. In particular, the two-dimensional spin-1/2 Ising model and other exactly solvable models play an important role in statistical mechanics of cooperative phenomena. Although, in most of theoretical studies the exchange interaction is usually assumed to be of short-range, in many real magnetic insulators the exchange coupling between magnetic atoms is predominantly carried out by means of the indirect exchange through the intermediary non-magnetic atom (the so-called superexchange interaction). In order to describe these materials theoretically, Fisher introduced a two-dimensional antiferromagnetic Ising model on the decorated square lattice, in which the interaction between decorating magnetic atoms realized via intermediate non-magnetic atoms was considered \cite{Fisher1,Fisher2}. 

It is also well known that the Fisher superexchange antiferromagnet is one of the few exactly solvable models in the presence of an external magnetic field. In fact, until its establishing, the Ising model in a nonzero magnetic field was included in the class of unsolved problems in statistical mechanics.
Consequently, the effect of external magnetic field on magnetic properties was investigated only applying various approximate methods, the solutions of which were in many cases in no good agreement with each other as well as with experiments. The typical example of the experimental as well as theoretical discussions
is the behavior of initial susceptibility in antiferromagnetic solids. Some experiments and approximations predicted that susceptibility has a maximum at a critical point and falls away on either its site. According to others, the  zero-field susceptibility passes through a maximum up to above the critical temperature. At the critical temperature it shows the infinite slope.

Moreover, no approximate methods yield explicit expressions for the initial susceptibility as a function of temperature, nor do they indicate how it behaves in a finite magnetic field. The only case in which the magnetic susceptibility as well as other thermodynamic quantities can be exactly calculated is Fisher two-dimensional model. However, this model is only one of a number of similar exactly solvable antiferromagnetic lattices and its simple generalization leads to a series of other models which could display a variety of interesting features. To the best of our knowledge, only two generalizations of Fisher model have been investigated so far, namely, the superexchange Ising model on the kagome lattice \cite{Wu} and the decorated Ising model which is the combination of the Fisher model and the Shiozi-Miyazima model of dilute-Ising spin system \cite{Mashiyama}. 

Considering this fact, the main purpose of the present work is to generalize the original superexchange model for arbitrary values of decorating atoms including also the effect of the uniaxial crystal-field anisotropy as well.

The outline of this paper is as follows. In the next section the basic points of the exact solution are briefly  presented. In Sec.3 we discuss the most interesting results for the ground-state phase diagram, critical temperature  and thermodynamics properties of the system under investigation. Finally, some concluding remarks are given in Sec.4.

\section{Model and method}
In this article we will consider a generalized Fisher superexchange antiferromagnet on the square lattice (see Fig.1), in which vertex sites are occupied by the fixed spin-$1/2$ atoms (open circles) and decorating sites by the spin-$S$ atoms ($S > 1/2$) (closed and obliquely lined circles). Hence, we assume that horizontal and vertical decorating spins interact with each other through the so-called superexchange leading to an antiferromagnetic long-range order. Similarly to Fisher \cite{Fisher1,Fisher2}, we also suppose that all horizontal decorating spins are coupled with their nearest neighbors through an antiferromagnetic interaction, whereas in the vertical direction, there is a ferromagnetic nearest-neighbor exchange interaction.

Furthermore, including the effects of external longitudinal field and uniaxial crystal-field anisotropy 
on all decorating atoms, the total Hamiltonian of the system can be written in the form
\begin{eqnarray}
\label{eq:Hd}
\hat{{\cal{H}}}_d 
&=& 
J \sum_{\langle i,j \rangle}^{2N} \hat{S}_{i}^{z}\hat{\mu}_{j}^{z} 
- J \sum_{\langle k,l \rangle}^{2N} \hat{S}_{k}^{z}\hat{\mu}_{l}^{z}
 - D\sum_{i \in B}^{2N} (\hat{S}_{i}^{z})^2 \ - H\sum_{i \in B}^{2N} \hat{S}_{i}^{z},
\end{eqnarray}

\begin{wrapfigure}{i}{0.5\textwidth}
\centerline{\includegraphics[width=0.46\textwidth]{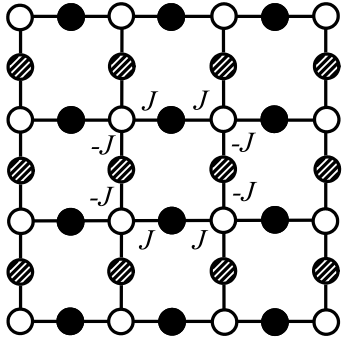}}
\caption{ Part of the decorated square lattice with vertex spin 1/2 (open circles) and decorating spins 3/2 (closed
and obliquely lined circles. J denotes the exchange interaction which is antiferromagnetic (ferromagnetic) in the horizontal (vertical) direction}
\label{fig1}
\end{wrapfigure}
where $\hat{\mu}_{j}^{z}$ and $\hat{S}_{i}^{z}$, respectively, denote the standard spin operators of spin-1/2 and spin-$S$ Ising atoms, the first and second terms describe the exchange interaction between the nearest neighbors in the horizontal and vertical direction and the last two expressions describe the interaction of decorating atoms with  the uniaxial crystal-field anisotropy $D$ and the longitudinal magnetic field $H$, respectively. Finally, the parameter $J>0$ represents the exchange integral between the nearest neighbors.

For later convenience, it is useful to rewrite the total Hamiltonian (\ref{eq:Hd}) 
in the form $\hat{{\cal{H}}}_d = \sum_{h = 1}^N \hat{{\cal{H}}}_h + \sum_{v = 1}^N \hat{{\cal{H}}}_v$, 
where the first (second) term represents the sum of bond Hamiltonians involving all interaction terms 
associated with the decorating atoms on the horizontal (vertical) bonds and that are defined as follows:
\begin{eqnarray}
\label{eq:HhHv}
\hat{{\cal{H}}}_k &=& \alpha J \hat{S}_{k}^{z} ( \hat{\mu}_{k1}^{z} + \hat{\mu}_{k2}^{z} ) - \hat{S}_{k}^{z} H  - 
D (\hat{S}_{k}^{z})^2  
\end{eqnarray} 
with $k = h$ and $v$ for horizontal and vertical bonds, respectively. The parameter $\alpha$ specifies the type of exchange interaction, thus for $k = h $ we put $\alpha= 1$ (antiferromagnetic exchange) and for $k = v $ is $\alpha= -1$ (ferromagnetic exchange).

The most important point of our calculation is the evaluation of the partition function ${\cal{Z}}_{d}$ of 
the investigated system. The validity of commutation relation for the bond Hamiltonians (i.e. $[\hat{{\cal{H}}}_i, \hat{{\cal{H}}}_{j}] = 0$,  for  $i \neq j$) enables one to rewrite ${\cal{Z}}_{d}$ in the partially factorized form, namely
\begin{eqnarray}
\label{eq:Zd}
{\cal {Z}}_{d} = 
\mathrm{Tr}\exp (-\beta \hat{{\cal{H}}}_{d}) = 
\mathrm{Tr}_{\{\mu\}}\prod_{\langle h,v \rangle }^{N} {\cal{Z}}_{h}\,{\cal{Z}}_{v}.
\end{eqnarray}
In the above, $\beta = 1/k_{B}T$, ($k_{B}$ being Boltzmann constant and $T$ the absolute temperature) and $\mathrm{Tr}_{\{\mu\}}$ means a trace over the degrees of freedom of spin-1/2 Ising vertex spins. Finally, the bond partition functions ${\cal{Z}}_{h}$ and ${\cal{Z}}_{v}$ furnish traces over all remaining degrees of freedom and they are given by
\begin{equation}
\label{eq:ZhZv}
{\cal{Z}}_{k} = \mathrm{Tr}_{S_{k}}\,\exp (-\beta\hat{{\cal{H}}}_{k}) =
\sum_{n = - S}^{S} 
\exp(\beta D n^2)\,\cosh\Big(\beta n\,[\,J (\hat{\mu}_{k1}^{z} + \hat{\mu}_{k2}^{z}) - \alpha H\,] \Big) \,, 
\end{equation}       
where $\mathrm{Tr}_{S_{k}}$ denotes a trace over spin-$S$ decorating atom on the $k$th horizontal ($k = h$) or vertical ($k = v$) bond, respectively. The latter relation implies the possibility of introducing the 
the decoration-iteration mapping transformation \cite{Syozi,Fisher3} 
\begin{eqnarray}
\label{eq:DIT}
\mathrm{Tr}_{S_{k}}\exp(-\beta\hat{{\cal{H}}}_{k}) 
= 
A \exp\,\big(\,\beta R\hat{\mu}_{k1}^{z}\hat{\mu}_{k2}^{z} + \beta H_{0k}(\hat{\mu}_{k1}^{z} + \hat{\mu}_{k2}^{z})/4\,\big)\, \quad k = h, v.
\end{eqnarray}
Considering all possible configurations of the spins $\hat{\mu}_{k1}^{z}$ and $\hat{\mu}_{k2}^{z}$ in previous  
equations one finds that $H_{0h} = -H_{0v}$. Consequently, when equation (\ref{eq:DIT}) is substituted into (\ref{eq:Zd}), the magnetic contributions $H_{0h}$ and $H_{0v}$ belonging to the vertex spins of the lattice cancel out and the partition function of the system ${\cal{Z}}_{d}$ reduces to the form
\begin{equation} 
\label{eq:ZdZ0}
{\cal{Z}}_{d}(\beta, J , D, H) = A^{2N} {\cal{Z}}_{0}(\beta, R) .
\end{equation}
Here, ${\cal{Z}}_{0}$ represents the partition function of the standard spin-$1/2$ Ising square lattice without external longitudinal magnetic field and the transformation  parameters $A$ and $R$ are given by
\begin{eqnarray}  
\label{eq:AR}
A = \{W(J)W(-J)W(0)^{2}\}^{1/4}\,\,\,\,, 
\qquad \beta R = \ln\Bigg\{\frac{W(J) W(-J)}{W(0)^{2}}\Bigg\}\,, 
\end{eqnarray}  
where $W(x)$ depends on the temperature, external magnetic field and spin $S$ of decorating atoms and it is defined as 
\begin{equation}  
\label{eq:W1W2W3}
W(x) = \sum_{n = -S}^{S}\exp(\beta D n^2) \cosh (\beta n x - \beta n H ) . 
\end{equation}
It is worth noticing that equation (\ref{eq:ZdZ0}) relates the partition function of the studied model in the presence of an external magnetic field and that of the standard zero-field spin-$1/2$ Ising model on the square lattice. Since the explicit expression for ${\cal{Z}}_{0}$ is  known \cite{Onsager}, we can straightforwardly calculate many relevant physical quantities based on the familiar thermodynamic relations. On the other hand, some important quantities, such as the staggered magnetization or correlation functions cannot be obtained within thermodynamic approach. Fortunately, this problem can be solved utilizing the following exact spin identities \cite{Strecka}
\begin{eqnarray}
\langle
f_{1}(\hat{\mu}_{i}^{z},\hat{\mu}_{j}^{z},...,\hat{\mu}_{k}^{z})  
\rangle_{d} 
&=& 
\langle
f_{1}(\hat{\mu}_{i}^{z},\hat{\mu}_{j}^{z},...,\hat{\mu}_{k}^{z})
\rangle_{0} \,,  \label{eq:f1} \\
\langle
f_{2}(\hat{S}_{k}^{z},\hat{\mu}_{k1}^{z},\hat{\mu}_{k2}^{z})    
\rangle_{d} 
&=& 
\Bigg \langle
\frac{\mathrm{Tr}_{S_{k}} f_{2}(\hat{S}_{k}^{z},\hat{\mu}_{k1}^{z},\hat{\mu}_{k2}^{z})
\mathrm{exp}(-\beta\hat{{\cal{H}}}_{k})}
{\mathrm{Tr}_{S_{k}}\mathrm{exp}(-\beta\hat{{\mathcal{H}}}_{k})}
\Bigg\rangle_{d} \,,\qquad  k = h,v.  
\label{eq:f1f2}
\end{eqnarray}
where arbitrary function $f_{1}$ depends on vertex spin variables, the function $f_{2}$ depends on the spin variables of the $k$th bond only and the symbols $\langle ... \rangle_{0}$ and $\langle ... \rangle_{d}$ stand for the standard canonical averages of the original and decorated lattice, respectively. For example, applying the identity (\ref{eq:f1f2}), one simply attains the following results for the sublattice magnetization in the horizontal direction
\begin{equation}
     m_{Bh}^z \equiv \langle \hat{S}_{k}^{z} \rangle_d = A_0 + A_1 m_A^z + A_2 \varepsilon_A,  
 \label{mbh}
 \end{equation} 
where coefficients $A_i$ depend on the temperature, external magnetic field and crystal field. Moreover, we have introduced the sublattice magnetization $m_A^z \equiv \langle \hat{\mu}_{i}^{z} \rangle_{d}$ 
and correlation function $\varepsilon_A \equiv \langle \hat{\mu}_{i1}^{z}\hat{\mu}_{i2}^{z} \rangle_{d}$. Very similar equations but with different coefficient can also be  derived for the correlation function $C_{k}^z \equiv \langle \hat{\mu}_{k}^{z}\hat{S}_{k}^{z} \rangle_{d}$ and quadrupolar momentum $q_{k}^z \equiv \langle (\hat{S}_{k}^{z})^2 \rangle_d$ with $k = h, v$. Although the derivation of final equations for the above-mentioned quantities is straightforward, the calculation by itself is lengthy and tedious. Therefore we do not present details here.

Finally, let us briefly comment on the choice of the order parameter in our model. In general, at low temperatures the antiferromagnets are distinguished by the long-range order with anti-parallel alignment of the nearest-neighboring magnetic moments (or spins). This type of magnetic arrangement is usually described by the so-called staggered magnetization which is defined as a difference of reduced sublattice magnetization. It is also well-known that Ising-type antiferromagnets in higher dimensions undergo a second-order phase transition at some critical temperature and the staggered magnetization is then nonzero below the critical temperature and vanishes above this temperature.

However, as we have already mentioned at the beginning of this section, the long-range antiferromagnetic order between pairs of horizontal and vertical decorating atoms of the lattice is realized indirectly through intermediate non-magnetic vertex atoms. Moreover, the vertex atoms (that are not coupled to the external magnetic field) also exhibit a spontaneous magnetization below the transition temperature. For this purpose, it is convenient to define the order parameter of the model under investigation in the form of binary vector \mbox{${\bf m} = ( |m_B^s|, m_A^z)$}, where the first vector component represents the absolute value of staggered magnetization ($m_B^s = (m_{Bh}^z - m_{Bv}^z)/2$) describing the long-range antiferromagnetic ordering of horizontal and vertical decorating atoms, while the second one describes the spontaneous magnetization of the spin-$1/2$ Ising sublattice. 

\section{Numerical results and discussion}
Before discussing the most interesting numerical results, it is worth emphasizing that the superexchange model is generally slightly different from the standard antiferromagnetic Ising models on the square lattice usually treated by most authors. In particular, the standard antiferromagnetic model becomes ferromagnetic when the sign of exchange integral $J$ is changed, whereas the superexchange model remains invariant against the transformation $J \rightarrow -J$. It is also worth noting that for $S=1/2$ and $S=1$ our calculation recovers the results obtained by Fisher,   and Mashiyama and Nara \cite{Fisher1,Fisher2,Mashiyama}. For this reason we concentrate here on the case $S=3/2$ which has not been discussed yet. 
\begin{table}[h]
\caption{Values of sublattice magnetization, correlation functions and quadrupolar momentum for different ground-state phases.}
\label{tbl-smp1}
\vspace{2ex}
\begin{center}
\renewcommand{\arraystretch}{0}
\begin{tabular}{|c||c|c|c|c|c|c|c|c|}
\hline
& $m_A^z$ &  $m_{Bh}^z$ & $m_{Bv}^z$ &$\varepsilon_A $ & $C_{h}^z $ & $C_{v}^z $ & $q_{h}^z$ & $q_{v}^z$ \strut\\
\hline
\rule{0pt}{2pt}&&&&&&&&\\
\hline
    SAP$_1$ & 0.5 & -1.5 &  1.5 &  0.25 &  -0.75 &  0.75&  2.25 & 2.25  \strut\\
\cline{1-9}
    SAP$_2$ & 0.5   &  -0.5   &  0.5 &  0.25  &  -0.25     &  0.25   &  0.25 &  0.25   \strut\\
\hline
   SAP & 0.5 &  -0.5 & 1.5 & 0.25 & -0.25 & 0.75  &  0.25 &  2.25  \strut\\
\cline{1-9}
 FIFP  & 0.5 &  0.5 &  1.5 &  0.25 &  0.25 &  0.75 & 0.25 & 2.25 \strut\\
\hline
 PP$_1$& 0.0 & 1.5 & 1.5 &  0.0 & 0.0 & 0.0 &  2.25 & 2.25 \strut\\
\cline{1-9}
 PP$_2$ & 0.0 &  0.5 &  0.5  & 0.0 &  0.0 &  0.0 & 0.25 &  0.25\strut\\
\hline
\end{tabular}
\renewcommand{\arraystretch}{0}
\end{center}
\end{table}
 
Now, let us proceed with the discussion of the ground state. Possible phases at $T=0.0$ can be identified throught the analysis of all sublattice magnetization, pair-correlation functions and quadrupolar momenta belonging to the decorating spins and those of original undecorated Ising lattice. The results of our investigation are collected in table 1. As we can see, there exist six different phases, namely three superantiferromagnetic phases (SAP$_1$, SAP$_2$ and SAP), one field-induced ferromagnetic phase (FIFP) and two paramagnetic phases (PP$_1$ and PP$_2$). For a better illustration, we have depicted in figure 2 the ground-state phase diagram in the $D - H$ plane.
\begin{wrapfigure}{i}{0.5\textwidth}
\centerline{\includegraphics[width=0.46\textwidth]{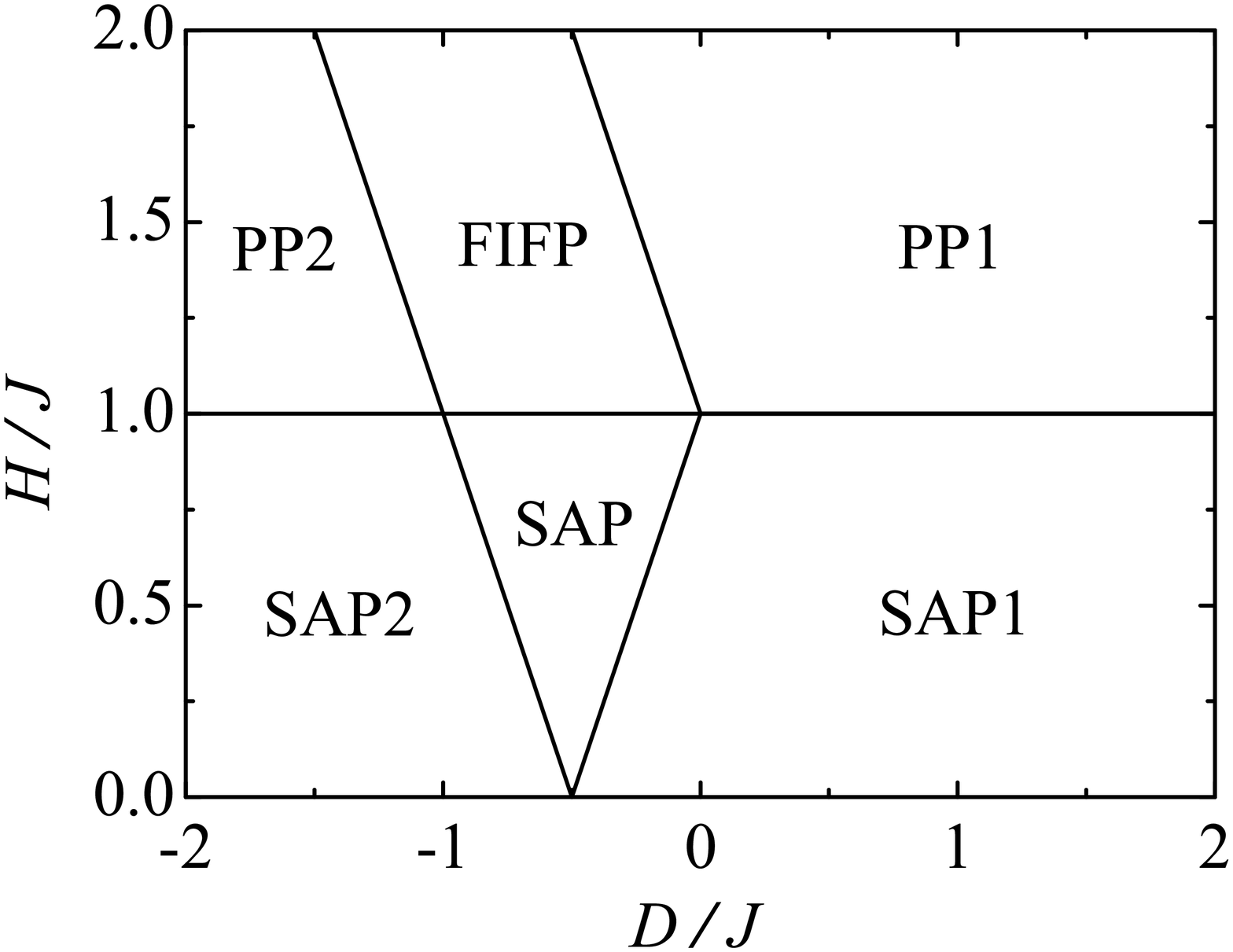}}
\caption{Grond-state phase diagram in the $D-H$ plane for the decorated superexchange antiferromagnetic model
with $S = 3/2$.}
\label{fig2}
\end{wrapfigure}
Comparing the quantities evaluated above one easily observes that for sufficiently small magnetic fields, the decorating sublattice of the system always exhibits an antiferromagnetic long-range order. This is characterized by the mutually opposite signs of sublattice magnetization belonging to them: $m_{Bh}^z = -3/2$ and $m_{Bv}^z = 3/2$ in the SAP$_1$, \mbox{$m_{Bh}^z = -1/2$ and $m_{Bv}^z = 1/2$} in the SAP$_2$ and $m_{Bh}^z = -1/2$ and $m_{Bv}^z = 3/2$ in the SAP. Moreover, the sublattice magnetization $m_A^z$ takes its saturated value in the whole region where the superantiferromagnetic phases are stable. Consequently, the spin-1/2 Ising atoms localized in vertexes of the lattice exhibit here exclusively the perfect ferromagnetic arrangement.

On the other hand, the magnetic field stronger than a certain critical value ($H_c/J \ge 1.0$) overturns anti-parallel decorating spins to its direction and consequently destroys the long-range antiferromagnetic ordering in the system. Therefore, in the region of magnetic fields $H > H_c$, both sublattice magnetization $m_{Bh}^z$ and $m_{Bv}^z$ will be positive: $m_{Bh}^z = m_{Bv}^z = 3/2$ in the PP$_1$, $m_{Bh}^z = m_{Bv}^z = 1/2$ in the PP$_2$ and $m_{Bh}^z = 1/2$, $m_{Bv}^z = 3/2$ in the FIFP. In fact, for $H>H_c$ the FIFP is the only stable ordered ground-state phase, which always exists in a relatively narrow region of the crystal fields. It is also interesting, that in spite of the external magnetic
\begin{wrapfigure}{i}{0.5\textwidth}
\centerline{\includegraphics[width=0.45\textwidth]{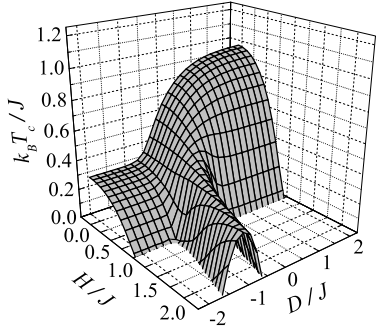}}
\caption{Global phase diagram in the $D-H-T$ space for the decorated superexchange antiferromagnetic model
with $S = 3/2$.}
\label{fig3}
\end{wrapfigure}
field being applied only on the decorating atoms, its sufficiently strong value indirectly effects the alignment of the Ising spins in the vertexes of lattice. In fact, in both paramagnetic phases PP$_1$ and PP$_2$ the vertex non-magnetic spins are strongly frustrated due to the competitive effect of the magnetic field, crystal field and  antiferromagnetic exchange interaction. Indeed, the zero values of all relevant correlation functions (see table 1) clearly confirm the presence of frustration. One also finds that  at the exact phase boundaries, the relevant phases co-exist implying the possibility of the first-order phase transition at $T=0.0$. Thus the ground-state behavior of our model is much more complex in comparison with the original Fisher model.

Apart from the ground state properties, we have also calculated the critical temperatures at which the system undergoes the second-order phase transition. Our results for the global phase diagram in the $D-H-T$ space are shown in figure 3. As we can see, the critical temperature exhibits a very interesting behavior. Naturally, the ordered (disordered) phases are stable bellow (above) the critical temperature. Of course, in some regions the stability of disordered phases extends down to zero temperature in agreement with the ground-state phase diagram. One should 
\begin{wrapfigure}{i}{0.5\textwidth}
\centerline{\includegraphics[width=0.45\textwidth]{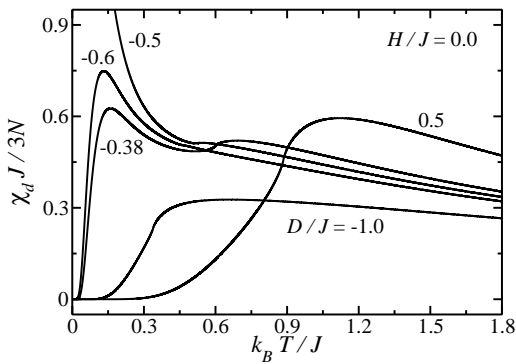}}
\caption{Temperature dependences of the reduced initial susceptibility for the decorated superexchange antiferromagnetic model with $S = 3/2$ and some typical values of the crystal field.}
\label{fig4}
\end{wrapfigure}
also note that our model exhibits essentially the same critical behavior as Fisher and Mishiyama and Nara models \cite{Fisher1,Fisher2,Mashiyama}. This is obvious, since the decoration of the lattice by finite number of atoms cannot lead to a change of the universality class of the model. As we have already mentioned above, one of the most interesting physical quantities in antiferromagnetic models represents the initial suceptibilty. In order to understand the behavior of this quantity, in figure 4 we have presented thermal variation of the initial susceptibility (i.e $H=0$) for several typical values of the crystal-field anisotropy. Besides the standard curves
\begin{figure}[t]
\centerline{\includegraphics[width=0.85\textwidth]{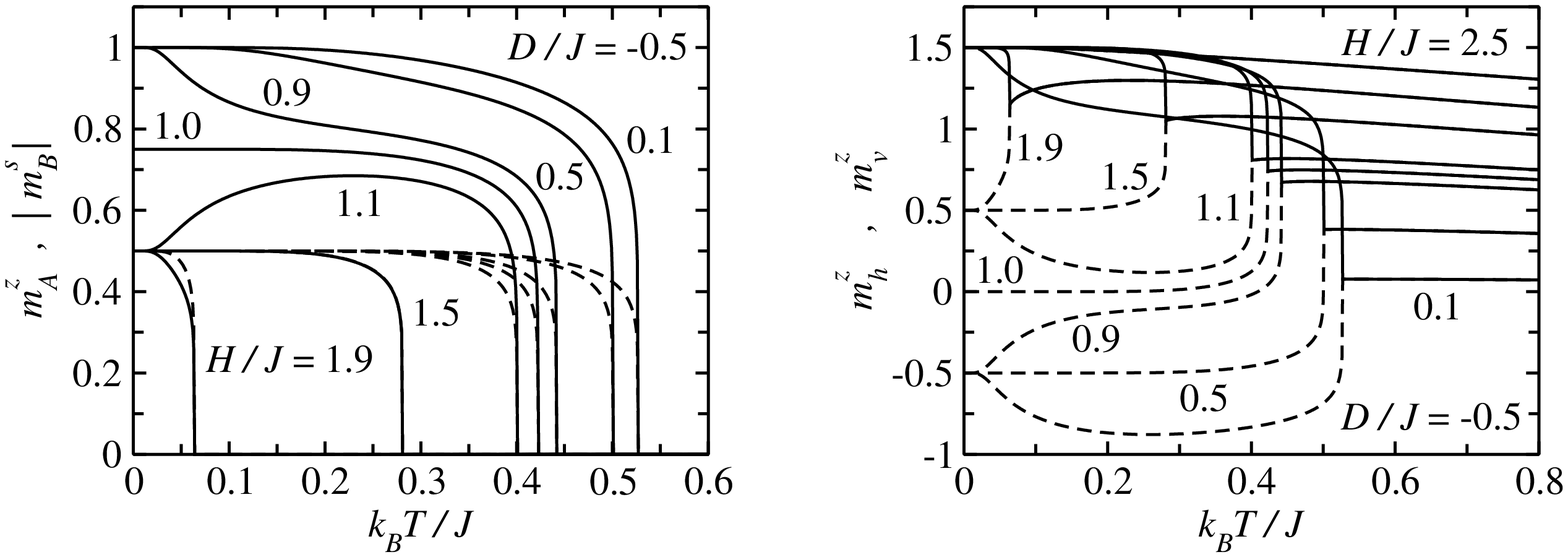}}
\caption{Temperature dependences of the reduced  components $m^s_B, \; m_A^z$ of the order-parameter and the reduced sublattice magnetization $m_{Bh}^z, \; m_{Bv}^z$ for the decorated superexchange antiferromagnetic model with $S = 3/2, \; D/J = -0.5$ and some typical values of the magnetic field.}
\label{fig5}
\vspace*{-0.55mm}
\end{figure}
(see the curve labeled  $D/J = 0.5$), we also obtained some new dependences with clear maxima in the low-temperature region (see curves labeled $D/J= -0.6$ and $-0.38$). Moreover, one finds that strong nagative value of the crystal field rapidly depresses the absolute value of the initial susceptibility and causes its very slow decreasing above $T_c$ (see the case $D/J = -1.0$). This kind of dependences cannot be observed in the original Fisher model, since they originate from the crystal-field effects. The curve obtained for $D/J = -0.5$ also represents a furter original result which exhibits a strong divergence as temperature approaches zero. This behavior indicates the appearence of the first-order phase transition, since in this case we are exactly located at the phase boundary of the ground state.

\begin{figure}[h]
\centerline{\includegraphics[width=0.85\textwidth]{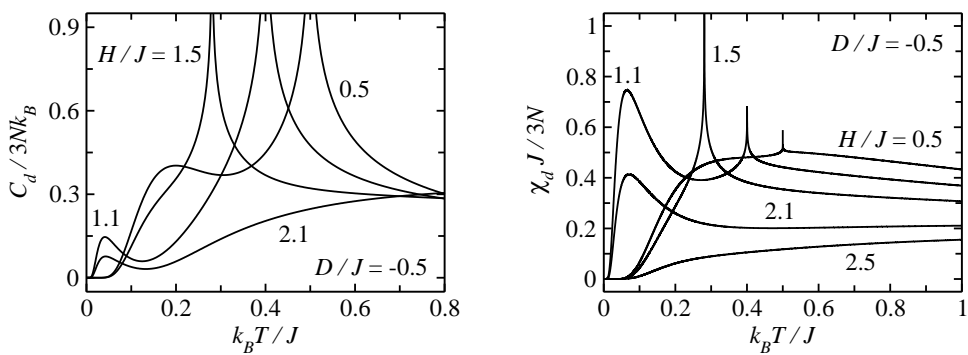}}
\caption{Temperature dependences of the reduced specific heat and susceptibility for the decorated superexchange antiferromagnetic model with $S = 3/2, \; D/J = -0.5$ and some typical values of the magnetic  field.}
\label{fig6}
\end{figure} 
Finally, to illustrate the behavior of other physical quantities at finite temperatures, in figures 5 we have depicted thermal variations of the order parameter components $m^s_B, \; m_A^z$ and magnetization $m_{Bh}^z, \; m_{Bv}^z$ of decorating spins. In both figures we have fixed $D/J = -0.5$ and have changed the values of magnetic field. Comparing the behavior of the relevant quantities in figure 5 one observes the characteristic behavior of the order parameter components being nonzero below $T_c$ and vanishing at all temperatures above $T_c$. On the other hand $m_{Bh}^z, \; m_{Bv}^z$ exhibit nonzero values in the whole temperature region. Thus they cannot serve as good order parameters for the model. At $T=0.0$ all quantities take the corresponding characteristic values depending on the strength of the magnetic field. Of particular interest is the curve for $H/J =1.0$ which apparently indicates the co-existemce of the SAP$_1$ and FIFP at $T=0.0$ in agreement with the ground-state phase diagram. The existence of different phases and competitive effect of the magnetic field, crystal-field anisotropy and temperature are also reflected in the thermal dependences of the specific heat and susceptibility that are shown in figure 6. Both quantities exhibit characteristic behavior with interesting low-temperature maxima and logarithmic divergence at the critical temperature.

\section{Conclusion}
In this work we have generalized the Fisher superexchange antiferromagnetic model for the case of arbitrary decorating spins. Excepting the external magnetic field we have also investigated the effect of the uniaxial crystal-field anisotropy. Some preliminary results for the case of $S=3/2$ decorating atoms have revealed very rich and interesting behavior. Further interesting results for the magnetization process, thermal and field dependences of the entropy, specific heat and susceptibility will be presented in the near future. Moreover, more realistic versions of this model including the transverse magnetic field or biaxial crystal-field anisotropy are also exactly solvable and may potentially enrich our knowledge about exactly solvable models in statistical mechanics.

\section*{Acknowledgments}
This work was financially supported under grants VEGA 1/2009/05 and APVT 20-005204 %

\end{document}